\documentclass[onecolumn]{elsart}
\usepackage{amssymb}
\usepackage{epsfig}
\journal{Physics Letters A}
\begin{document}

\begin{frontmatter}

\title{Stability of spinning ring solitons of the cubic-quintic nonlinear
Schr\"{o}dinger equation}

\author{Isaac Towers $^1$}
\author{Alexander V. Buryak $^1$},
\author{Rowland A. Sammut $^1$},
\author{Boris A. Malomed $^2$},
\author{Lucian-Cornel Crasovan $^3$} and
\author{Dumitru Mihalache $^3$}

\address{$^1$ School of Mathematics and Statistics,\\ University of New South Wales at ADFA,
 Canberra, ACT 2600, Australia}
\address{$^3$ Department of Interdisciplinary Sciences, Faculty of\\
Engineering, Tel Aviv University, Tel Aviv 69978, Israel}
\address{$^3$ Department of Theoretical Physics, Institute of Atomic\\
Physics, P.O. Box MG-6, Bucharest, Romania}
\begin{abstract}
We investigate stability of (2+1)-dimensional ring solitons of the
nonlinear Schr\"{o}dinger equation with focusing cubic and
defocusing quintic nonlinearities. Computing eigenvalues of the
linearised equation, we show that rings with spin (topological
charge) $s=1$ and $s=2$ are linearly stable, provided that they
are very broad. The stability regions occupy, respectively, $9\%$
and $8\%$ of the corresponding existence regions. These results
finally resolve a controversial stability issue for this class of
models.
\end{abstract}

\end{frontmatter}

\section{Introduction}

Recently, much interest has been focused on {\it spinning} optical
solitons, i.e., those carrying topological charge, in both (2+1)D
\cite{unstable1,DVS,experiment,Q,tristram} and (3+1)D
\cite{dark,Anton,new2,new3} geometries. A spinning soliton has an
embedded phase dislocation and carries intrinsic angular momentum.
The integer number of phase rotations around the dislocation is
the soliton's topological charge or ``spin''.

Broadly speaking, spinning solitons can be divided into two
classes: (i) dark, i.e., vortices produced by a phase dislocation
which is embedded in an infinite background; and (ii) bright, with
the vortex core embedded in a bright (localised) multidimensional
soliton proper; the amplitude of which vanishes at infinity. In
this Letter we consider bright spinning solitons in the (2+1)D
geometry. In terms of nonlinear optics, (2+1)D solitons may be
naturally realised as spatial solitons in the form of cylindrical
beams in a bulk medium, or, alternatively, as spatiotemporal
solitons in the form of fully localised ``light bullets''
propagating in a planar waveguide (film). Due to the presence of
the vorticity, the soliton's cross section has an annular shape.
We shall refer to them in what follows below as {\it localised
vortex solitons} (LVSs). As for the (3+1)D solitons, they are
spatiotemporal ``bullets'' propagating in bulk media.

Models that may give rise to stable multidimensional, (2+1)D or
(3+1)D, solitons must necessarily have nonlinearity which prevents
dynamical collapse. Well-known examples of collapse-free
nonlinearities are $\chi ^{(2)}$ (second-harmonic-generating),
saturable, and cubic-quintic (CQ); the cubic and quintic
components being, respectively, self-focusing and self-defocusing.
All these nonlinearities occur in various optical media.


However, unlike ground-state (zero-spin) (2+1)D bright solitons,
for the spinning ones the absence of collapse does not guarantee
dynamical stability. In fact, a LVS tends to be strongly
destabilised by azimuthal perturbations which break it up into
several separating zero-spin bright solitons (the latter are
stable). In (2+1)D models with $\chi ^{(2)}$ and saturable
nonlinearities, numerical simulations had revealed strong
azimuthal instability \cite{unstable1,DVS}, which was later
observed experimentally in a $\chi ^{(2)}$ medium
\cite{experiment}. A similar instability of (3+1)D rings resulting
from the $\chi ^{(2)}$ nonlinearity has been found in Ref.
\cite{new3}: a 3D LVS (in fact, it is a torus) breaks up due to an
azimuthal perturbation, and the resulting zero-spin solitons fly
off in directions tangential to the ring.

In terms of the LVS stability, more promising are models with
competing nonlinearities. The first step in this direction was the
study of (2+1)D rings in the CQ nonlinear Schr\"{o}dinger
equation. Simulations reported by Quiroga-Teixeiro and Michinel
\cite{Q} showed that broad rings found (by means of the
variational approximation and direct numerical methods) in that
model not only did not demonstrate growth of small perturbations,
but also survived collisions, thus appearing fairly stable.
However, a similar analysis for the (3+1)D CQ model \cite{new2}
has demonstrated that narrow LVSs were broken up quickly, while
broad rings were destabilised much slower by azimuthal
perturbations. However, eventually all the LVSs for which a
definite numerical result could be obtained were unstable. This
suggests re-checking the above-mentioned stability of LVSs in the
2D model reported in \cite{Q}. Rerunning simulations for the same
cases which were considered in \cite{Q}, it was established
\cite{PhysicaD} that, in fact, they are {\em also subject} to the
weak instability against azimuthal perturbations, provided that
the simulations are long enough.

As 2D and 3D LVSs in the CQ model become very broad, it still
remains unknown whether the growth rate of their instability
against azimuthal perturbations gradually vanishes in the limit of
infinitely broad rings, or if there is a clearly defined
transition to truly stable solitons. The problem of discerning
between very weak instability and true stability may be of little
importance for applications, as experiments are always carried out
in finite samples, which do not have enough room for the
development of an instability if it is extremely weak.
Nevertheless, the issue is of principal interest and is therefore
worthy of consideration.

Another model with competing nonlinearities which may be promising
for the generation of stable rings combines the $\chi ^{(2)}$ and
{\em self-defocusing} cubic [$\chi ^{(3)}$] nonlinearities. It is
necessary to say that no conventional nonlinear material with
strong $\chi ^{(2)}$ nonlinearity directly satisfies the
requirement of this model to have a negative $\chi ^{(3)}$
coefficient at both the fundamental- and second-harmonic
frequencies (see below). However, two possibilities to create the
necessary {\em effective} $\chi ^{(3)}$ nonlinearity have been
proposed: (i) by creating a layered medium in which layers
providing for the $\chi ^{(2)}$ nonlinearity periodically
alternate with others that account for the self-defocusing Kerr
nonlinearity, and (ii) by engineering special $\chi ^{(2)}$
quasi-phase-matched gratings \cite{Bang}. In the latter case,
induced $\chi ^{(3)}$ and intrinsic $\chi ^{(2)}$ nonlinearities
may be equal in strength, and the former one may be given either
sign.

Recently, LVSs were considered in this $\chi ^{(2)}-\chi ^{(3)}$
model \cite{isaac}. Using direct simulations and linear stability
analysis, it has been shown that narrow rings demonstrate typical
breakup into zero-spin bright solitons initiated by the azimuthal
instability, but very broad (flat-top) LVSs with the spin
(topological charge) $s=1$ and $s=2$ were indeed found to be {\em
completely stable}. In these two cases, the stability region is,
respectively, $\approx 8\%$ and $\approx 5\%$ of the corresponding
existence domain.

In the {\it cascading limit}, corresponding to large wave-vector
mismatch, the $\chi ^{(2)}-\chi ^{(3)}$ model reduces to the CQ
model, which suggests that there may be a chance to find
completely stable LVSs in the CQ model too, which is a subject of
the present work. We will employ the same techniques that were
applied in Ref. \cite{isaac} to the $\chi ^{(2)}-\chi ^{(3)}$
model, i.e., rigourous computation of the stability eigenvalues.
This will overcome limitations of ``phenomenological'' analyses of
the CQ model carried out in previous works, which relied on
simulations of perturbed solitons to directly test their
stability, or simulations of the linearised equation to estimate
the largest growth rate of instability. A shortcoming of both
methods is that they can miss very weak instability. We will
demonstrate that LVSs in the CQ model are {\em truly} stable if
their width (or energy) exceeds a certain critical value. This
result completes the investigation of the CQ model in the (2+1)D
case, which was the subject of many above-mentioned works, and
suggests a challenging question: can spinning ``bullets'' with
sufficiently large energy be stable in the same model in the
(3+1)D case?

\begin{figure}\epsfxsize=100mm \centerline{\epsfbox{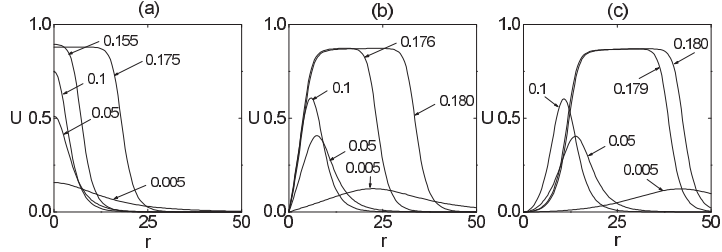}}
\caption{The stationary (2+1)D spinning-soliton solutions: (a)
$s=0$, (b) $s=1$, and (c) $s=2$. The values of $\protect\kappa $
are indicated near the curves.} \label{profiles}
\end{figure}

\section{The ring solitons}

Following the derivation of Ref. \cite{Anton}, but assuming two
transverse spatial dimensions, we arrive at the dimensionless CQ
equation
\begin{equation}
i\frac{\partial u}{\partial z}+\nabla _{\perp
}^{2}u+|u|^{2}\,u\,-|u|^{4}\,u\,=0.  \label{scaled}
\end{equation}
where $u$ is the envelope of the electromagnetic wave propagating
along the $z$-direction in the optical medium. In the case of the
cylindrical beams in the bulk medium (i.e., spatial solitons, see
above), the Laplacian in Eq. (\ref{scaled}) is the diffraction
operator acting on the transverse spatial coordinates $x$ and $y$.

In the alternative case of spatiotemporal solitons in the planar
waveguide, $y$ is replaced by a properly scaled temporal variable,
$\tau \equiv t-z/V_{0} $, where $t$ is time, and $V_{0}$ is the
group velocity of the carrier wave. In the latter case, the part
$\partial ^{2}/\partial \tau ^{2}$ in the Laplacian accounts for
the temporal dispersion (which must be {\it anomalous}, in order
to have to right sign), rather than diffraction, and LVS will be,
in unrescaled coordinates, a compressed (elliptic) ring moving in
its own plane.

Ring solitons are localised stationary solutions to Eq.
(\ref{scaled}) of the form
\begin{equation}
u=U(r)\exp (is\theta +i\kappa z),  \label{U}
\end{equation}
where $r$ and $\theta $ are polar coordinates in the ($x,y$)
plane, $\kappa $ is a wave number shift (relative to the carrier
wave), and the integer $s$ is the spin. The amplitude $U$ can be
taken to be real, obeying an equation
\begin{equation}
\frac{\partial ^{2}U}{\partial r^{2}}+\frac{1}{r}\frac{\partial
U}{\partial r}-\frac{s^{2}}{r^{2}}U-\kappa U+U^{3}\,-U^{5}\,=0.
\label{stationary}
\end{equation}

We solved Eqs. (\ref{stationary}) by means of the relaxation
technique. When $\kappa $ is small (a low-power regime), LVSs are
narrow. The beam's amplitude at first increases with $\kappa $ and
then saturates, while the ring's width keeps increasing because of
the self-defocusing effect of the cubic term.

The wave number $\kappa $ parameterises a family of stationary
solutions, examples of which are displayed in Fig. \ref{profiles}.
The existence regions for solitons in the 2D and 3D versions of
the CQ model are known \cite{Q,Anton}:
\begin{equation}
0<\kappa <\kappa _{{\rm offset}}^{{\rm (2D)}}\approx 0.18;0<\kappa
<\kappa _{{\rm offset}}^{{\rm (3D)}}\approx 0.15,\, \label{offset}
\end{equation}
and in both cases they practically do not depend on the soliton's
spin \cite {Anton}. The width and energy of LVS diverge as $\kappa
\rightarrow \kappa _{{\rm offset}}$. Note that a soliton solution
to the 1D version of Eq. (\ref {scaled}) is known in an exact
elementary form, the corresponding exact offset wave number being
$\kappa _{{\rm offset}}^{{\rm (1D)}}=3/16\equiv 0.1875$, so that
the values (\ref{offset}) are close to it.

\section{Stability}

As a precursor to the full linear stability analysis of the ring
solitons, we first consider the modulational stability of plane
continuous-wave (cw) solutions to Eq. (\ref{scaled}). This may
give some insight into the stability of broad rings as they tend
to these cw solutions in the limit $\kappa \rightarrow \kappa
_{{\rm offset}}$. The cw solutions with the propagation constant
$\kappa $ are
\begin{equation}
u_{0}=a_{0}\exp (i\kappa z),\,a_{0}^{2}=\frac{1}{2}\left( 1\pm
\sqrt{ 1-4\kappa }\right) .  \label{cw}
\end{equation}
We take small perturbations to the plane wave of the form
\begin{equation}
u_{1}=\left[ \alpha \exp \left( ikx+i\Omega z\right) +\beta \exp
\left( -ikx-i\Omega ^{\ast }z\right) \right] \exp (i\kappa z),
\label{u1}
\end{equation}
where $k$ is an arbitrary perturbation wave number, and $\Omega $
is the stability eigenvalue (the asterisk stands for the complex
conjugation); the cw solutions being stable if $\Omega $ is real
for all real $k$. Substituting into Eq. (\ref{scaled}) the
perturbed solution $u=u_{0}+u_{1}$ and linearising around $u_{0}$,
we find after some straightforward algebra that:
\begin{equation}
\Omega ^{2}=k^{2}+k^{4}\pm k^{2}\sqrt{1-4\kappa }-4k^{2}\kappa .
\end{equation}

It is easy to demonstrate that for the cw branch with the upper
sign in Eq. (\ref{cw}) (i.e., with the larger amplitude), $\Omega
^{2}\geq 0$ for all $k$ , hence, this branch is always stable,
while the other one is not. As it was said above, the LVS
solutions of the CQ model that we are dealing with tend to the
stable cw solution as $\kappa \rightarrow \kappa _{{\rm offset}}$,
giving an initial indication that, as the rings broaden, they may
become stable.

\begin{figure}\epsfxsize=100mm \centerline{\epsfbox{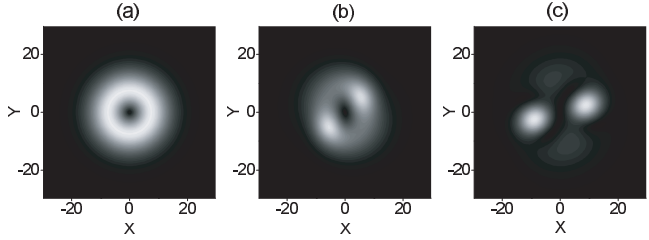}}
\caption{Gray-scale contour plots illustrating the evolution of an
unstable ring soliton with $s=1$ and $\protect\kappa=0.05$: (a)
$z=0$; (b) $z=310$; (c) $z=330$.} \label{propagate}
\end{figure}

This conclusion is suggested by simulations reported in Ref.
\cite{Q}, where sufficiently broad rings appeared to be stable,
whereas narrower ones were definitely unstable against azimuthal
perturbations (see Fig. \ref{propagate} ). On the other hand, as
was mentioned above, more accurate (longer) simulations of the
cases that were considered in that work show that weak instability
still occurs. To perform the direct simulations we used the
Crank-Nicholson scheme as a finite-difference approximation to
propagation equations (\ref{scaled}). The corresponding system of
nonlinear equations was solved by means of the Picard iteration
method ( see details in Ref. \cite{Otega}). Typically, we chose
equal transverse grid sizes $\Delta x=\Delta y=0.4$ and the
longitudinal grid size was $\Delta z=0.02$. Thus, direct
simulations show a general trend of suppression of the azimuthal
instability with the increase of the size of LVSs. This may be
sufficient to expect experimental observability of LVSs, but,
following this analysis, one cannot predict if the instability may
be completely eliminated, provided that the size of LVS exceeds
some critical value.

The stability issue can only be resolved in the precise sense by
comprehensive analysis of eigenmodes of small perturbations around
LVS. To this end, we add infinitesimal complex perturbations
$\epsilon (z,r,\theta )$ to the stationary solutions of Eqs.
(\ref{scaled}) and (\ref{stationary}) with the vorticity $s$, cf.
Eq. (\ref{u1}),

\begin{equation}
u=[U(r)+\epsilon (z,r,\theta )]\exp (is\theta +i\kappa z).
\label{stat}
\end{equation}
A general perturbation $\epsilon (z,r,\theta )$ may always be
expanded into a series, with each term having its own vorticity
$J$, so that a generic independent perturbation term is
\begin{equation}
\epsilon =\xi _{J}^{+}(r)\exp \left( i\lambda z+iJ\theta \right)
+\xi _{J}^{-}(r)\exp \left( -i\lambda ^{\ast }z-J\theta \right) ,
\label{perturb}
\end{equation}
where $\lambda $ is the (generally complex) LVS's instability
eigenvalue. Substituting this into Eqs. (\ref{scaled}) and
linearising, we arrive at a non-self-adjoint eigenvalue problem:
\begin{equation}
{ \lambda \vec{\xi _{J}}=\left[
\begin{array}{cc}
C_{+} & D \\
-D & -C_{-}
\end{array}
\right] \vec{\xi _{J}},}  \label{eigen}
\end{equation}
where $\hat{\xi _{J}}\equiv (\xi _{J}^{+},\xi _{J}^{-})$, $C_{\pm
}=\hat{L}_{J}^{\pm }+2U^{2}-3U^{4}$, $D=U^{2}-2U^{4}$, and
\[
\hat{L}_{J}^{\pm }\equiv \frac{\partial ^{2}}{\partial
r^{2}}+\frac{1}{r}\frac{\partial }{\partial
r}-\frac{1}{r^{2}}(s\pm J)^{2}-\kappa .
\]

Instability is accounted for by eigenvalues with ${\rm Im} \lambda
\neq 0$ (the present system being Hamiltonian, eigenvalues appear
in complex conjugate pairs or quadruplets). The continuous
spectrum of the eigenvalues consists of real intervals $\kappa
\leq \lambda <\infty $ and $-\infty <\lambda \leq -\kappa $.

To analyse the eigenvalue problem (\ref{eigen}), we replaced the
differential operators by their fifth-order finite-difference
approximations and solved the resulting algebraic eigenvalue
problem numerically. We mostly used grids with $400$ to $800$
points, but up to $1200$ points were used in regions where a
change of the stability occurs. To verify the precision of the
numerical code, we also used another technique based on the
relaxation method for solving two-point boundary-value problems.
Although limited to finding real eigenvalues, the latter method
admits a high degree of precision control without much of the
computational overhead of other methods. For instance, it has been
recently used to a great effect in finding a small stability
window for higher-order solitons in a third-harmonic-generation
model, which would have otherwise been overlooked \cite{Kazimir}.
Comparison between the spectral and relaxation methods has shown
that the former one has good precision for the number of the grid
points used: the numerical error in calculating the
stability-boundary values of $\kappa$ is estimated to be $\delta
\kappa_{{\rm st}}\sim 10^{-5}$ for $1200$ grid points.

\begin{figure}\epsfxsize=70mm \centerline{\epsfbox{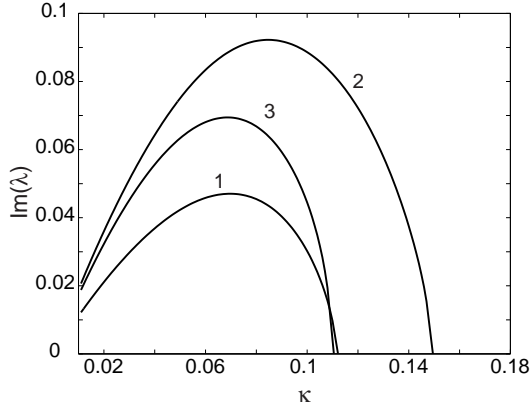}}
\caption{Unstable eigenvalues for the ring solitons with $s=1$.
The spin $J$ of the azimuthal perturbation is indicated next to
each curve (the perturbations with $J>3$ caused no instability,
therefore they are not displayed). Only Im$(\protect\lambda )$ is
shown. Note that the dominant instability has $J=2$, and it
vanishes at $\protect\kappa \approx 0.16$, while the ring solitons
exist up to $\protect\kappa =\protect\kappa _{{\rm offset}}\approx
0.18$.} \label{charge1}
\end{figure}

\begin{figure}\epsfxsize=70mm \centerline{\epsfbox{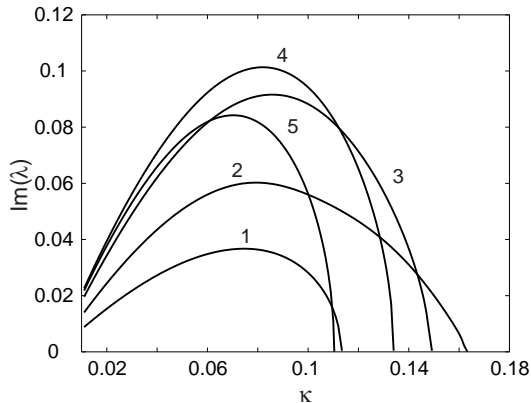}}
\caption{The same as in the previous figure but for $s=2$.}
\label{charge2}
\end{figure}

Results of the linear stability analysis for the fundamental
($s=1$) and next-order ($s=2$) LVSs are summarised in Figs.
\ref{charge1} and \ref{charge2}. We considered perturbations with
$J=0,...,\pm 5$, and have found that instability of the
fundamental LVSs is not generated by perturbations with $J>3$. The
most persistent instability mode in the case $s=1$ corresponds to
$J=\pm 2$. Subsequent direct simulations demonstrate that this
instability mode, if it takes place, initiates eventual breakup of
the ring into several zero-spin solitons. In Ref. \cite{DVS} an
estimate for the number of filaments $N$ resulting from the
destruction of the LVS  is given as $N\approx 2s$. Our results
agree with this estimate: for LVSs with $s = 1$, dominant
instability corresponds to $J = \pm 2$ and a singly charged LVS
breaks into two filaments, as can be seen in Fig. \ref{propagate}.
(This agreement largely holds for the $s=2$ case as well where for
most of the unstable domain the dominant instability corresponds
to $J=\pm4$ and typically doubly charged LVSs break into four
filaments.)

In all the cases considered, we have found that there is a {\em
stability-change} value $\kappa _{{\rm st}}$, at which the largest
instability eigenvalue ${\rm Im}\lambda $ vanishes, and remains,
along with all the other ones, exactly (up to the numerical
accuracy) equal to zero in the {\em stability window}, $\kappa
_{{\rm st}}<\kappa <$ $\kappa _{{\rm offset}}$ (recall $\kappa
_{{\rm offset}}$ is the upper existence boundary of the LVS
family); in other words, thin LVSs are unstable and broad ones are
stable. The existence of the window is clearly illustrated by Fig.
\ref {charge1}. For the fundamental LVSs ($s=1$), the stability
window occupies $\approx 9\%$ of the existence domain $[0,\kappa
_{{\rm offset}}]$.

For the LVSs with $s=2$, a similar situation takes place. The
dominant instability in a larger part of their existence domain is
generated by the perturbation with $J=4$, but for broad LVSs it is
overtaken by the $J=2$ mode. The linear spectrum of the LVSs with
$s=2$ contains a stability window which occupies $\approx 8\%$ of
the existence region.

It appears that stable LVSs cannot have the value of the spin
larger than $2$. In particular, the LVSs with $s=3$ were found to
demonstrate a persistent weak instability associated with the
$J=\pm 1$ perturbation modes at all the values of $\kappa $, and
it seems very plausible that higher-order LVSs will continue to do
so.

In the work of Quiroga-Teixeiro and Michinel \cite{Q}, the authors
use variational techniques to make an estimate for the width of
the stability window. Using the formulas derived in Ref. \cite{Q}
the ratio of $\kappa_{\rm st}/\kappa_{\rm offset}$ for $s=1$ and
$s=2$ LVSs is 0.803 and 0.838 respectively. In other words
according to the variational analysis of \cite{Q} the stability
window occupies approximately 20\% and 16\% of the existence
domain $[0,\kappa _{{\rm offset}}]$ for singly and doubly charged
LVS. These estimates correctly predict the existence of stability
domains for both $s=1$ and $s=2$ LVSs and also the decrease in
stability window size for the doubly charged LVSs relative to the
singly charged. The factor of 2 difference between the estimates
and the numerical results obtained in this work is quite
reasonable considering the ansatz used in Ref. \cite{Q} becomes
increasingly less accurate for large $\kappa$ i.e. where the
stable LVSs exist.

Lastly, it is relevant to mention that, in the work
\cite{PhysicaD}, a variational approach was used to study, in an
analytical form, instability of LVSs in the (2+1)D and (3+1)D
versions of the CQ model against a special perturbation mode in
the form of an infinitesimal shift in the position of the vortex
core relative to the broad soliton as a whole. In terms of the
expansion (\ref{perturb}), this mode corresponds to $J=\pm 1$. If
$\delta H$ is a small variation of the value of the soliton's
Hamiltonian generated by the infinitesimal off-centre shift of the
vortex core, an instability condition which is well-known from the
general soliton stability theory is $ \delta H<0$
\cite{stability}. Assuming $\kappa $ close enough to $\kappa_{{\rm
offset}}$, which is necessary to have broad solitons whose outer
radius is much larger than the radius of the vortex core, it has
been shown that a shift of the core leads to a {\em decrease} of
an effective Hamiltonian of the interaction between the core and
the outer rim of the soliton, which implies an instability.
Because the interaction Hamiltonian is exponentially small in the
case of the broad LVS, the instability is also expected to be
exponentially weak. This instability mode was predicted for $s=1$
and $2$, but not for $s\geq 3$. It is noteworthy too that only for
$s=1$ the thus predicted instability is linear (exponentially
growing), while for $s=2$ it is nonlinear ({\em sub}exponential).

\begin{figure}\epsfxsize=70mm \centerline{\epsfbox{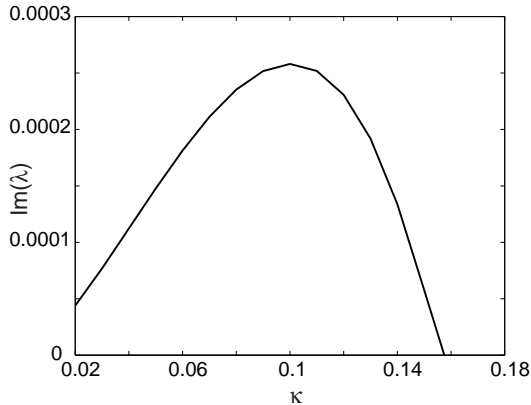}}
\caption{Weak instability of the ring solitons with $s=1$
resulting from $J=\pm1$ perturbations. To the limit of the
numerical method, this mode has zero instability growth rate at
$\protect\kappa \geq 0.16$. A similar mode can be found for the
$s=2$ rings.} \label{bubble}
\end{figure}

From our numerical results, we have indeed found a weak
instability mode with $J=1$ for the $s=1$ and $s=2$ rings, see
Fig. \ref{bubble}. To the limit of our numerical accuracy, the
corresponding instability growth rate ${\rm Im}\lambda $ vanishes
at $\kappa \approx 0.16$, and then remains equal to zero up to the
point $\kappa =\kappa _{{\rm offset}}^{{\rm (2D)}}\approx 0.18$,
at which the size of the ring becomes infinitely large, see Eq.
(\ref {offset}). It is not completely clear yet whether an
extremely small exponentially vanishing unstable eigenvalue exists
past the value $\kappa \approx 0.16$, but the issue is purely
formal, the broad rings being stable in any practical sense.

In conclusion, we have found that localised vortex solitons, with
values of the spin $1$ and $2$, of the two-dimensional
cubic-quintic nonlinear Schr\"{o}dinger equation are linearly
stable in a finite interval of the propagation constant
corresponding to broad solitons with large power. These results
make an important step forward in the resolution of the
controversial stability issue in this class of models.

I.T., A.V.B., and R.A.S. acknowledge support from the Australian
Research Council. B.A.M. appreciates support from the Binational
(US-Israel) Science Foundation.

\end{document}